\documentclass[prl,superscriptaddress,preprint,endfloats,nopacs]{revtex4-2}

\usepackage{amsmath, amsthm, amssymb, pifont, wasysym}

\usepackage{graphicx}
\usepackage{bm} 
\usepackage{dcolumn} 
\usepackage{color}
\usepackage[version=4]{mhchem}
\usepackage{gensymb}
\definecolor{mygray}{gray}{0.5}

\newcommand {\EAX}{\ce{Eu$A$2$X$2}}
\begin{document}

\title{\ce{EuIn2Sb2}: epitaxially stabilized axion insulator candidate with strong spin-orbit coupling}

\author{Hsiang Lee}
\affiliation{Department of Physics, Institute of Science Tokyo, Tokyo 152-8551, Japan}
\author{Shinichi Nishihaya}
\affiliation{Department of Physics, Institute of Science Tokyo, Tokyo 152-8551, Japan}
\author{Markus Kriener}
\affiliation{RIKEN Center for Emergent Matter Science (CEMS), Wako 351-0198, Japan}
\author{Ayano Nakamura}
\affiliation{Department of Physics, Institute of Science Tokyo, Tokyo 152-8551, Japan}
\author{Tadashi Yoneda}
\affiliation{Department of Physics, Institute of Science Tokyo, Tokyo 152-8551, Japan}
\author{Yuki Deguchi}
\affiliation{Department of Physics, Institute of Science Tokyo, Tokyo 152-8551, Japan}
\author{Makuro Goto}
\affiliation{Department of Physics, Institute of Science Tokyo, Tokyo 152-8551, Japan}
\author{Masaki Uchida}
\email[Author to whom correspondence should be addressed: ]{m.uchida@phys.sci.isct.ac.jp}
\affiliation{Department of Physics, Institute of Science Tokyo, Tokyo 152-8551, Japan}
\affiliation{Toyota Physical and Chemical Research Institute, Nagakute 480-1192, Japan}

\begin{abstract}

Eu triangular lattice layer compounds described by the general formula $\EAX$ have attracted growing attention due to a wide range of layered crystal structures and corresponding magnetic topological phases, arising from the coexistence of large Eu magnetic moments and energetically inverted $A$ and $X$ bands. 
Among the $\EAX$ family, \ce{EuIn2Sb2} has been predicted to host robust axion insulator and higher-order topological insulator states by stronger spin-orbit coupling but has not been experimentally realized, including the structural identification. Here we report the epitaxial stabilization of \ce{EuIn2Sb2} by adopting the molecular beam epitaxy technique. Structural characterization reveals that \ce{EuIn2Sb2} films are based on a unique In-on-In stacking, forming a new trilayer structure distinct from previously predicted structures in the $\EAX$ family. 
In addition to the in-plane antiferromagnetic ordering at $T_\text{N}=10.5$~K, the stronger spin-orbit coupling and enhanced In--In hybridization in \ce{EuIn2Sb2} make it an ideal candidate of axion insulator and higher-order topological insulator phases. These results establish \ce{EuIn2Sb2} as a new member of the $\EAX$ family and offer an advanced platform for exploring these magnetic topological phases.

\end{abstract}

\maketitle
\newpage

There has been growing interest in materials with nontrivial electronic band topology due to their symmetry-protected boundary states and unconventional transport properties \cite{hasan2010colloquium,qi2011topological,fu2007topological,fu2007topologicalIS,zhang2009topological,yan2012topological,wieder2022topological}. Especially when long-range magnetic order is present, the topological classification of electronic bands is further altered \cite{hasan2010colloquium,tokura2019magnetic,li2019intrinsic,he2020mnbi2te4,otrokov2019prediction}. As a result, magnetic topological materials can host a variety of magnetotransport phenomena not accessible in nonmagnetic systems, including the intrinsic anomalous Hall effect and its quantization \cite{nagaosa2010anomalous,chang2013experimental,deng2020quantum,bernevig2022progress}, Weyl orbits involving surface Fermi arcs \cite{wan2011topological,potter2014quantum,xu2015discovery,yan2017topological,armitage2018weyl,nishihaya2019quantized,zhang2021cycling,nishihaya2021intrinsic}, higher-order topology and associated boundary states \cite{schindler2018higher,schindler2018higherTI,tokura2019magnetic,xu2019higher,noguchi2021evidence,puthiya2025eua2x2}, and axion electrodynamics \cite{wan2011topological,xu2019higher,zhang2019topological,liu2020robust,li2010dynamical,riberolles2021magnetic}. Compared with magnetically doped topological insulators and artificial heterostructures, intrinsically magnetic compounds provide more well-defined and diverse magnetic order, facilitating research on topological band structures beyond simple ferromagnetic cases. Therefore, the exploration of novel magnetic topological materials is crucial for advancing our understanding of magnetic topological phases \cite{deng2020quantum,li2019intrinsic,otrokov2019prediction,bernevig2022progress,li2019dirac}.

\begin{figure}
\includegraphics[width=\linewidth]{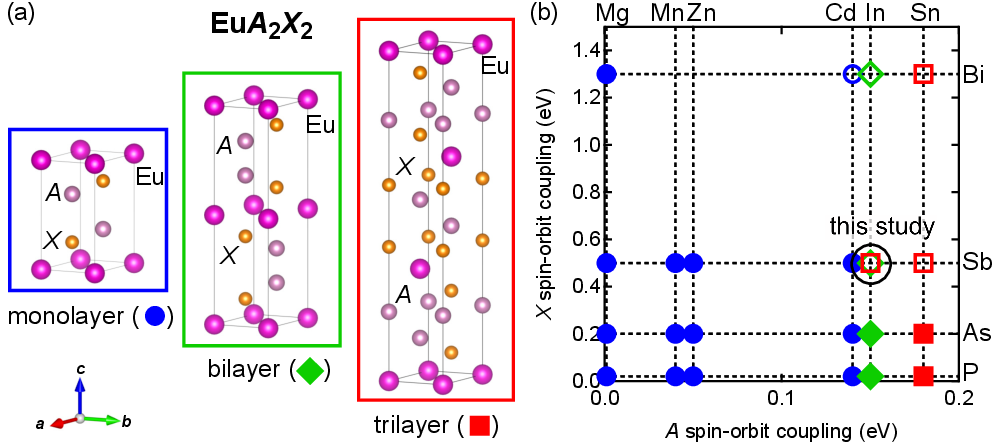}%
\caption{$\EAX$ compounds with Eu triangular lattice.
(a) Three different crystal structures of $\EAX$ ($A$ = Mg, Mn, Zn, Cd, and Sn; $X$ = P, As, Sb, and Bi), derived from the \ce{CaAl2Si2}-type monolayer structure.
(b) Experimentally synthesized or theoretically predicted $\EAX$ compounds with their crystal structures identified, mapped onto the spin-orbit coupling of $A$ and $X$ atoms. Here the spin-orbit coupling values are roughly estimated from atomic fine structure splittings based on NIST Atomic Spectra Database \cite{NIST_ASD}. Monolayer, bilayer, and trilayer structures with space groups $P\overline{3}m1$, $P6_3/mmc$, and $R\overline{3}m$ are denoted by a blue circle, a green diamond, and a red square, respectively. Solid symbols represent experimentally synthesized materials \cite{ma2019spin,nishihaya2024intrinsic,puthiya2025eua2x2,cuono2023ab,su2020magnetic,ohno2022maximizing,weber2006low,lee2025difference,arguilla2017eusn,goforth2008magnetic,gui2019new,wartenberg2002neue,payne2002synthesis,anand2016metallic,schellenberg2010121sb,may2011structure,berry2022type,goryunov2012esr,kranenberg2000structure,jiang2006colossal}, while open symbols correspond to those predicted only by calculations \cite{puthiya2025eua2x2,cuono2023ab,wang2021magnetic}. For \ce{EuIn2Sb2} studied in this work, two types of structures, bilayer and trilayer, have been predicted by first-principles calculations \cite{puthiya2025eua2x2,cuono2023ab}.}
\label{fig1}
\end{figure}

Magnetic compounds described by the general formula $\EAX$ with Eu triangular lattice layers constitute an important class of magnetic topological materials \cite{ma2019spin,soh2019ideal,su2020magnetic,puthiya2025eua2x2,xu2019higher,riberolles2021magnetic,gui2019new,pierantozzi2022evidence,yi2017large,marshall2021magnetic,heinrich2024magnetic,soh2026magnetic,weber2006low,cuono2023ab,pakhira2021type,payne2002synthesis,may2011structure,berry2022type,ohno2022maximizing,nakamura2024berry,nakamura2024plane,roychowdhury2023anomalous,li2019dirac,li2021magnetic,chen2020negative}. Here, the $A$ sites host Mg, Mn, Zn, Cd, In, and Sn atoms, while the $X$ sites host P, As, Sb, and Bi atoms. In some of the $\EAX$ compounds, particularly those with strong spin-orbit coupling, itinerant carriers originating from energetically inverted $A$ and $X$ bands coexist with large Eu$^{2+}$ ($4f^{7}$) spin magnetic moments, resulting in strong coupling between magnetic ordering and topological band structures. Depending on the combination  of the $A$ and $X$ atoms, $\EAX$ crystallizes in three different layered structures derived from the original \ce{CaAl2Si2}-type structure, as shown in Fig.~\ref{fig1}(a). Namely, monolayer, bilayer, and trilayer structures with different stacking patterns are stabilized, belonging to the space groups $P\overline{3}m1$, $P6_3/mmc$, and $R\overline{3}m$, respectively \cite{ma2019spin,nishihaya2024intrinsic,puthiya2025eua2x2,cuono2023ab,su2020magnetic,ohno2022maximizing,weber2006low,lee2025difference,arguilla2017eusn,goforth2008magnetic,gui2019new,wartenberg2002neue,payne2002synthesis,anand2016metallic,schellenberg2010121sb,may2011structure,berry2022type,goryunov2012esr,kranenberg2000structure,jiang2006colossal,nakamura2024plane}. The previously reported $\EAX$ compounds, along with their crystal structures, are summarized in Fig.~\ref{fig1}(b).

Concerning the magnetic ordering, many of the $\EAX$ compounds exhibit an A-type antiferromagnetic ground state, where ferromagnetically aligned Eu$^{2+}$ magnetic moments within the $ab$ plane are stacked antiferromagnetically along the $c$ axis \cite{heinrich2024magnetic,soh2026magnetic,weber2006low,soh2019ideal,ma2019spin,cuono2023ab,pakhira2021type,payne2002synthesis,may2011structure,berry2022type}, and can be readily driven into a forced ferromagnetic state under magnetic field. Because the electronic topology is highly sensitive to both crystal symmetry and magnetic ordering, this $\EAX$ family hosts a wide range of magnetic topological phases, including magnetic Weyl semimetals \cite{su2020magnetic,ohno2022maximizing,nakamura2024berry,ma2019spin,soh2019ideal,roychowdhury2023anomalous,nakamura2024plane}, magnetic topological insulators \cite{li2019dirac,li2021magnetic,chen2020negative,gui2019new,pierantozzi2022evidence,marshall2021magnetic}, axion insulators \cite{xu2019higher,riberolles2021magnetic,pierantozzi2022evidence}, and higher-order topological insulator candidates \cite{xu2019higher,puthiya2025eua2x2}. This diversity reflects the modulation of band topology through chemical substitution and magnetic tuning within a common structural motif.

Among the $\EAX$ family, \ce{EuIn2Sb2} is of particular interest because it has not yet been experimentally realized but is expected to host strong spin-orbit coupling favorable for topological band structures, as highlighted in Fig.~\ref{fig1}(b). First-principles calculations have predicted that \ce{EuIn2Sb2} is stabilized in either the bilayer ($P6_3/mmc$) \cite{puthiya2025eua2x2} or the trilayer ($R\overline{3}m$) structure \cite{cuono2023ab}. Importantly, both the bilayer and trilayer structures have been predicted to host symmetry-allowed axion insulator and higher-order topological insulator phases under antiferromagnetic ordering \cite{xu2019higher,riberolles2021magnetic,puthiya2025eua2x2,li2019dirac,pierantozzi2022evidence}. In this context, \ce{EuIn2Sb2} can be compared with the axion insulator candidate \ce{EuIn2As2}, which crystallizes in the bilayer structure within a unique In-on-In stacking \cite{xu2019higher,riberolles2021magnetic}, where In atoms are positioned directly above those below. \ce{EuIn2Sb2} can also be compared with the magnetic topological insulator candidate \ce{EuSn2As2}, which adopts the exfoliable trilayer structure \cite{li2019dirac,arguilla2017eusn,li2021magnetic,chen2020negative,pakhira2021type}. Since In and Sb are expected to host stronger spin-orbit coupling than Cd and As, respectively, \ce{EuIn2Sb2} lies in a significantly strong spin-orbit coupling regime, especially compared to the existing $\EAX$ systems, offering it as a promising platform for exploring magnetic topological phases.
\ce{EuIn2Sb2} is less stable than another Eu-In-Sb compound, \ce{Eu5In2Sb6}, which has been reported in bulk form \cite{park2002eu5in2sb6,rosa2020colossal}. In this case, molecular beam epitaxy (MBE) is a powerful technique for epitaxially stabilizing the \ce{EuIn2Sb2} phase.

In this work, we report the experimental realization of \ce{EuIn2Sb2} through epitaxial stabilization. High-quality \ce{EuIn2Sb2} thin films were grown by MBE, and their structural properties were systematically characterized. The films are found to adopt a novel trilayer structure based on the In-on-In stacking. Uniform stacking of the In-Sb-Eu-Sb-In repeating unit leads to a dominant trilayer structure, whereas alternating stacking results in the local formation of the original-predicted bilayer structure. 

\ce{EuIn2Sb2} thin films were grown by MBE on \ce{Al2O3} (0001) substrates in an EpiQuest RC1100 chamber \cite{nakazawa2019molecular,nakazawa2021enhancement,uchida2021above}. The \ce{Al2O3} substrate was preannealed to 700~$\degree$C to reconstruct the crystal surface.
 The substrate temperature was set to 550~$\degree$C during deposition, followed by annealing at the same temperature for approximately 30 hours. The beam equivalant pressures, measured by an ionization gauge, were $0.31\times10^{-5}$~Pa, $0.78\times 10^{-5}$~Pa, and $1.06\times10^{-5}$~Pa for Eu, In and Sb, respectively, corresponding to a flux ratio of 1:2.5:3.5. The film thickness was typically designed at about 70~nm, with a growth rate of $\sim 12$ \AA/min.
The crystal structure and phase purity were characterized by X-ray diffraction (XRD) and transmission electron microscopy (TEM) with energy-dispersive X-ray (EDX) spectroscopy mapping for Eu $L$, In $L$, and Sb $L$ edges. The out-of-plane interlayer spacing or lattice constant is determined by extrapolating the lattice parameters derived from a series of XRD $\theta-2\theta$ peak positions using the Nelson-Riley function. The in-plane lattice constants are calculated from the XRD reciprocal space mapping (RSM) peak positions.
Magnetotransport was measured by a standard four-probe method in a Cryomagnetics cryostat system equipped with a superconducting magnet. Magnetic properties were characterized using a superconducting quantum interference device magnetometer in a Quantum Design magnetic property measurement system (MPMS).

\begin{figure}
\includegraphics[width=\linewidth]{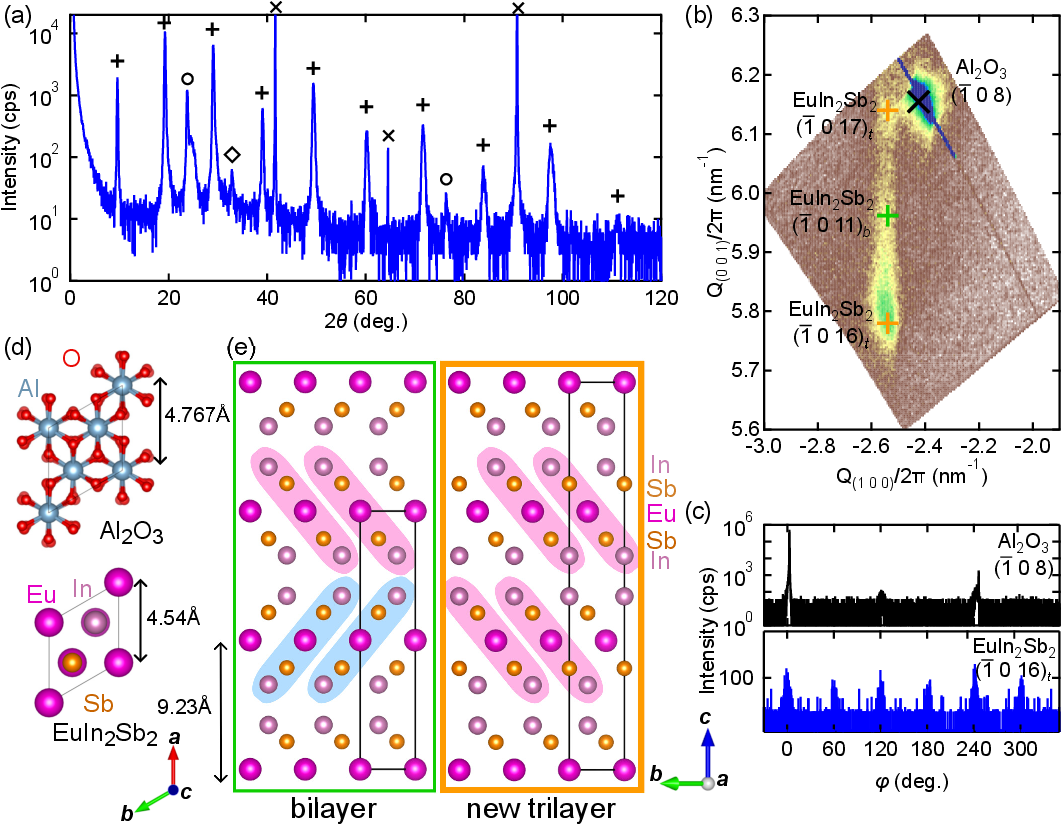}%
\caption{XRD characterization of \ce{EuIn2Sb2} film.
(a) XRD $\theta$-2$\theta$ scan of a \ce{EuIn2Sb2} thin film grown on an \ce{Al2O3} substrate. The series of sharp peaks marked by plus signs are indexed as the $(00\,2l)_b$ and $(00\,3l)_t$ peaks of \ce{EuIn2Sb2}, where $b$ and $t$ represent the bilayer and trilayer structures in (e), respectively. Substrate peaks are marked with crosses, and open circles and diamonds denote peaks originating from nonmagnetic impurity phases InSb and In, respectively.
(b) Reciprocal space map around the \ce{Al2O3} ($\overline{1}$ 0 8) peak.
(c) In-plane $\varphi$-scans of \ce{Al2O3} ($\overline{1}$ 0 8) and \ce{EuIn2Sb2} ($\overline{1}$ 0 16)$_t$ peaks.
(d) Planar epitaxial relation between \ce{Al2O3} and \ce{EuIn2Sb2}.
(e) Two types of crystal structures with In-on-In stacking, where In atoms are positioned directly above those below along the $c$ axis. These two structures are distinguished by alternating or uniform stacking of In-Sb-Eu-Sb-In unit layers. While the former corresponds to the bilayer structure in Fig. \ref{fig1}(a), the latter is a new trilayer structure, different from that in Fig.~\ref{fig1}(a) without the In-on-In stacking.}
\label{fig2} 
\end{figure}

Figure~\ref{fig2}(a) shows the XRD $2\theta$-$\omega$ scan of a \ce{EuIn2Sb2} thin film grown on an \ce{Al2O3} (0001) substrate. A series of sharp \ce{EuIn2Sb2} $(00l)$ peaks are clearly observed, indicating that the film is oriented along the $c$ axis. The weak impurity peaks mainly originate from a $\sim$10~nm InSb capping layer on the \ce{EuIn2Sb2} film, which was likely formed during the post-annealing process. It should be noted that the series of peaks observed in the $2\theta$-$\omega$ scan cannot distinguish between $(00l)_m$ of monolayer, $(002l)_b$ of bilayer structures, and $(003l)_t$ of trilayer structures; therefore, the number of Eu layers in the unit cell cannot be identified solely from this measurement. The out-of-plane Eu-Eu interlayer spacing along the $c$ axis is determined to be 9.23~\AA.

To examine the number of Eu layers together with the in-plane lattice constant, XRD RSM was performed. As shown in Fig.~\ref{fig2}(b), two clear peaks corresponding to the $(\overline{1}0\,16)_t$ and $(\overline{1}0\,17)_t$ reflections from a trilayer structure are observed, along with a weak $(\overline{1}0\,11)_b$ reflection from a bilayer structure. The in-plane lattice constants are determined to be $a=b=4.54$~\AA. Figure~\ref{fig2}(c) compares the $\varphi$ scans of \ce{Al2O3} $(\overline{1}08)$ and \ce{EuIn2Sb2} $(\overline{1}0\,16)_t$. The planar epitaxial relation between \ce{Al2O3} and \ce{EuIn2Sb2} is confirmed as sketched in Figs.~\ref{fig2}(d)-(e), and the sixfold symmetry observed for the \ce{EuIn2Sb2} peak indicates the presence of two domain types related by a 60$^\circ$ in-plane rotation in the trilayer structure.

We now turn to a detailed description of crystal structures characterized by the In-on-In stacking, expected for along the $c$ axis. As illustrated in Fig.~\ref{fig2}(d), alternating stacking of the repeating In-Sb-Eu-Sb-In unit layers gives rise to a bilayer ($P6_3/mmc$) structure, which is the same as that shown in Fig.~\ref{fig1}(a). In contrast, uniform stacking of the same unit layers may lead to another trilayer ($R\overline{3}m$) structure, which differs from the one shown in Fig.~\ref{fig1}(a). These structural models are consistent with the XRD results and are verified by the subsequent direct observations.

\begin{figure}
\includegraphics[width=\linewidth]{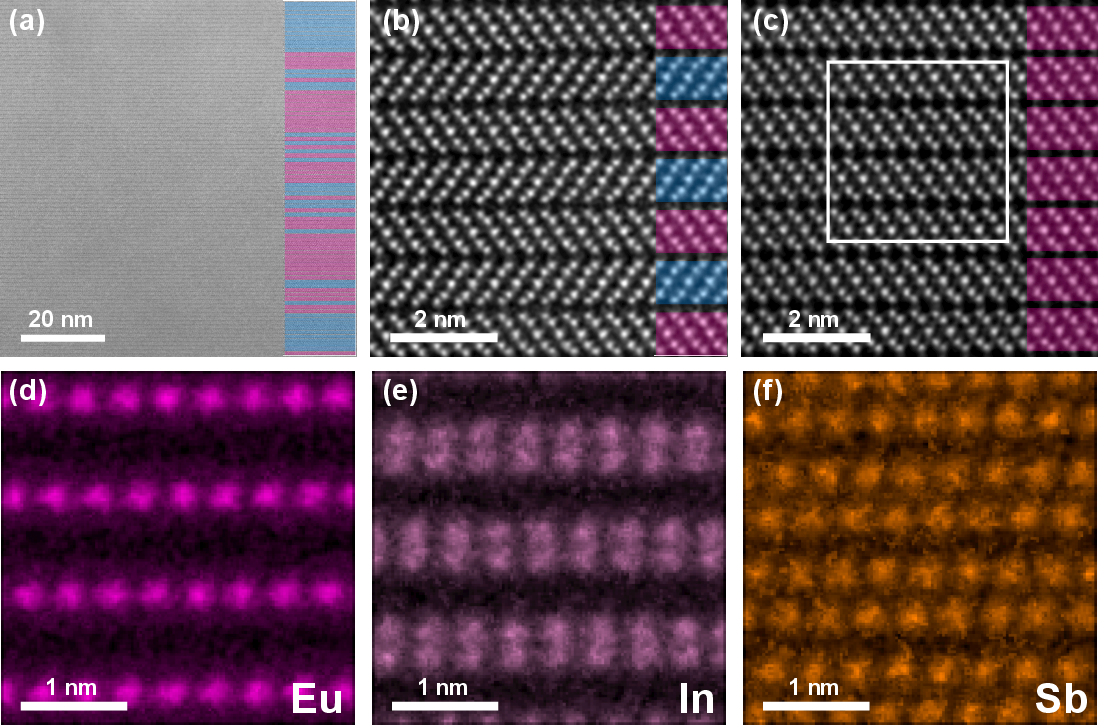}%
\caption{Atomically resolved cross-sectional structure images.
(a) Wide-area TEM image, and (b,c) higher-resolution HAADF-STEM images focusing on the bilayer and new trilayer structures, respectively. These are characterized by alternating or uniform stacking of the In-Sb-Eu-Sb-In layers, highlighted in red or blue according to their orientation.
EDX maps of the (d) Eu $L$, (e) In $L$, and (f) Sb $L$ edges in the boxed region in (c).}
\label{fig3} 
\end{figure}

To further examine the detailed crystal structure of \ce{EuIn2Sb2}, TEM observations were performed along with energy-dispersive X-ray (EDX) spectroscopy mapping. Fig.~\ref{fig3}(a) shows a wide-area TEM image of the \ce{EuIn2Sb2} film taken along the [100] direction of both \ce{EuIn2Sb2} and \ce{Al2O3}. While the characteristic In-on-In stacking is observed throughout the entire region, two different orientations of the In-Sb-Eu-Sb-In unit layers are identified, as highlighted by different colors. Alternating (AB-type) stacking of the unit layers, as magnified in Fig.~\ref{fig3}(b), occurs only in limited regions, resulting in the local formation of the bilayer structure shown in Fig.~\ref{fig2}(e). In contrast, the trilayer structure in Fig.~\ref{fig2}(e) dominates in the film, arising from uniform (ABC-type) stacking of the unit layers as shown in Fig.~\ref{fig3}(c) (See Supplemental Material for additional TEM images~\cite{supplemental}.) It should be emphasized that this trilayer structure is different from that originally predicted in Fig.~\ref{fig1}(a). It should be noted that the predicted bilayer structure has the same In-on-In configuration but a different stacking sequence, whereas the previously predicted trilayer structure has ABC-type stacking, in which the two In atoms between adjacent layers do not occupy the same in-plane site. Although both belong to the same space group $R\overline{3}m$, their stacking sequences are significantly different. As confirmed by the EDX mappings in Figs. \ref{fig3}(d)-(f), In atoms are stacked directly above those below, forming the In-on-In configuration, and the In-Sb-Eu-Sb-In unit layers exhibit ABC-type stacking as explicitly confirmed for Eu atoms.

\begin{figure}
\includegraphics[width=13cm]{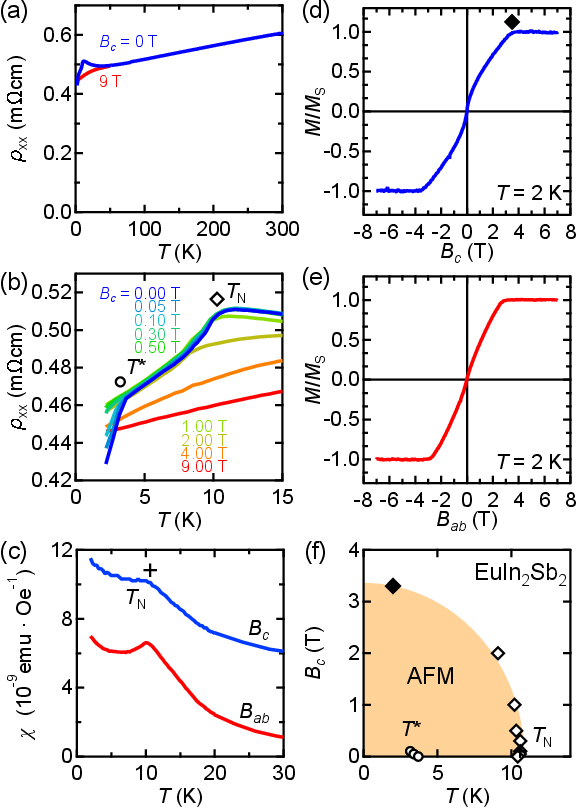}%
\centering
\caption{Magnetic phase diagram of \ce{EuIn2Sb2}.
(a) Temperature dependence of longitudinal resistivity $\rho_{xx}$ taken at out-of-plane magnetic fields $B_c$ of 0~T and 9~T.
(b) Expanded temperature dependence of $\rho_{xx}$ under various out-of-plane fields in finer steps. $T_\text{N}$ and $T^*$ are marked by an open diamond and circle, respectively.
(c) Temperature dependence of field-cooled magnetic susceptibility $\chi$ measured for out-of-plane field $B_{c}$ and in-plane field $B_{ab}$ of 0.1~T.
(d) Out-of-plane and (e) in-plane magnetization curves taken at 2~K. The magnetization is normalized to the saturation magnetization $M_s$ and the out-of-plane saturation field is marked by a diamond.
(f) Magnetic phase diagram of \ce{EuIn2Sb2} plotted for the $B_c - T$ plane, constructed from (b)-(d), where the open diamonds and circles show the $T_\text{N}$ and $T^*$, respectively, obtained from (b). The solid diamond is obtained from (d), and the cross is obtained from (c).
}
\label{fig4}
\end{figure}

The fundamental magnetotransport and magnetic properties of \ce{EuIn2Sb2} films are summarized in Fig.~\ref{fig4}. As shown in Fig.~\ref{fig4}(a), the longitudinal resistivity $\rho_{xx}$ exhibits a typical metallic temperature dependence and shows a clear kink at around 10.5~K. Although the metallic character varies among samples, this kink is reproducibly observed and is attributed to the antiferromagnetic transition temperature $T_\text{N}$ of the dominant trilayer \ce{EuIn2Sb2}. 

In the expanded view shown in Fig.~\ref{fig4}(b), another kink is observed at $T^* \sim 3.7$~K, in addition to that at $T_\text{N} = 10.5$~K. On the other hand, this anomaly at $T^*$ is absent in the subsequent magnetic susceptibility measurements. It is thus unlikely to originate from a further magnetic transition of \ce{EuIn2Sb2}, including the possibility of an antiferromagnetic transition of the local bilayer structure. On the other hand, note that in \ce{EuSn2As2}, a similar anomaly accompanied by a pronounced drop in resistivity, but not detected in careful thermodynamic measurements, has been observed at a temperature below $T_\text{N}$ \cite{pakhira2021type}. 
These arguments leave open the possibility that an additional intrinsic conduction path, such as edge or surface conduction, emerges below $T^*$, in these trilayer materials.

As shown in Fig.~\ref{fig4}(c), a kink corresponding to $T_\text{N} = 10.5$~K is observed in the magnetic susceptibility $\chi$ both for out-of-plane field $B_{c}$ along the $c$ axis and in-plane field $B_{ab}$ within the $ab$ plane. Especially for the out-of-plane field $B_{c}$, $\chi$ continues to increase significantly below $T_\text{N}$. This behavior is consistent with the magnetization curves measured for $B_c$ and $B_{ab}$ shown in Figs.~\ref{fig4}(d) and (e). For the out-of-plane field, the magnetization increases sharply with an S-shaped feature around zero field, even although the demagnetization effect in thin films is not taken into account in this plot. In contrast, the magnetization increases more gradually for the in-plane field $B_{ab}$. As summarized in Fig.~\ref{fig4}(f), the magnetic ground state of trilayer \ce{EuIn2Sb2} is based on the antiferromagnetic order with {Eu}$^{2+}$ spin magnetic moments lying in the $ab$ plane, possibly with a more complex modulation, such as the broken-helix spin structure observed in \ce{EuIn2As2} \cite{riberolles2021magnetic}.

We demonstrate that \ce{EuIn2Sb2} forms a new trilayer structure based on the unique In-on-In stacking configuration. In addition to the stronger spin-orbit coupling compared with \ce{EuSn2As2} and \ce{EuIn2As2}, this newly identified trilayer structure combines the symmetry of the \ce{EuSn2As2} trilayer structure with the In-In bonding configuration resembling that of the \ce{EuIn2As2} bilayer structure. In particular, the reduced In-In separation may enhance direct In-In hybridization, which is expected to increase the contribution of In $5s$ states near the Fermi level and thereby modify the low-energy electronic structure \cite{xu2019higher,sato2020signature,grigoriev2025universal,cuono2023ab}.

Under the in-plane antiferromagnetic ordering, the present trilayer structure as well as the bilayer structure breaks microscopic time-reversal symmetry while preserving inversion symmetry. These symmetry conditions are compatible with axion insulator and higher-order topological insulator phases \cite{xu2019higher,puthiya2025eua2x2,cuono2023ab,li2019dirac,pozo2019quantization,pierantozzi2022evidence}. In addition, the presence of mirror symmetries in these magnetic crystal structures can generate nonzero mirror Chern numbers, placing them in the class of topological crystalline insulator \cite{xu2019higher,li2019dirac,pierantozzi2022evidence}. Further optimization of growth conditions and elucidation of field-dependent magnetization ordering are key next steps. 

In summary, we have successfully demonstrated the epitaxial stabilization of the new compound \ce{EuIn2Sb2}. \ce{EuIn2Sb2} films are characterized by a unique In-on-In stacking configuration, forming a dominant trilayer structure belonging to the space group $R\overline{3}m$. This trilayer structure exhibits a stacking sequence distinct from the previously predicted one, adding a new structural member to the $\EAX$ family. The reduced In-In separation in the trilayer structure is expected to enhance direct In-In hybridization and influence the electronic states near the Fermi level. In addition to the in-plane antiferromagnetic ordering, the stronger spin-orbit coupling and the enhanced In-In hybridization in \ce{EuIn2Sb2} establish as an ideal candidate of axion insulator and higher-order topological insulator phases. Several future studies can build on the results of this work. First-principles calculations based on the newly observed trilayer structure and its experimentally determined lattice constants would help clarify its electronic structure. Further experiments could also investigate the predicted axion-related properties of \ce{EuIn2Sb2}. For example, angle-resolved photoemission spectroscopy (ARPES) could be used to examine the insulating surface states, transport measurements could test for zero Hall conductance, and optical measurements could probe the expected magneto-optical response. Our results provide an advanced platform for exploring these magnetic topological phases beyond the simple ferromagnetic case.

\begin{acknowledgments}

This work was supported by JSPS KAKENHI Grant Numbers JP23K13666, JP24H01614, JP24H01654, JP25H00841, and JP25K17447 from MEXT, Japan, by JST FOREST Program Grant Number JPMJFR202N and PRESTO Program Grant Number JPMJPR25H7, by Toyota Riken Rising Fellow Program funded by Toyota Physical and Chemical Research Institute, Japan, and by STAR Award funded by the Tokyo Tech Fund, Japan

The data are available from the authors upon reasonable request.
\end{acknowledgments}

%

\clearpage
\newpage

\clearpage
\newpage

\end{document}